\documentclass[twocolumn]{aastex61}
\usepackage{nicefrac}
\usepackage{graphicx}
\begin{document}
\bibliographystyle{apj}

\title{No Metallicity Correlation Associated with the Kepler Dichotomy}

\author{Carlos E. Munoz Romero}

\affil{Department of Physics, Grinnell College, 1116 8th Ave., Grinnell, IA 50112, USA}
\email{munozcar@grinnell.edu}

\author{Eliza M.-R. Kempton}

\affil{Department of Physics, Grinnell College, 1116 8th Ave., Grinnell, IA 50112, USA}
\email{kemptone@grinnell.edu}

\begin{abstract}

NASA's \textit{Kepler} mission has discovered thousands of planetary systems, $\sim 20\%$ of which are found to host multiple transiting planets. This relative paucity (compared to the high fraction of single transiting systems) is postulated to result from a distinction in the architecture between multi-transiting systems and those hosting a single transiting planet: a phenomenon usually referred to as the \textit{Kepler dichotomy}. In this paper, we investigate the hypothesis that external giant planets are the main cause behind the over-abundance of single- relative to multi-transiting systems, which would be signaled by higher metallicities in the former sample. To this end, we perform a statistical analysis on the stellar metallicity distribution with respect to planet multiplicity in the \textit{Kepler} data. We perform our analysis on a variety of samples taken from a population of $1166$ \textit{Kepler} main sequence planetary hosts, using precisely determined metallicities from the California-Kepler survey and \citet{swift}. Contrary to some predictions, we do not find a significant difference between the stellar metallicities of the single- and multiple-transiting planet systems.  However, we do find a $55\%$ upper bound for systems with a single non-giant planet that could also host a hidden giant planet, based on metallicity considerations. While the presence of external giant planets might be one factor behind the \textit{Kepler} dichotomy, our results also favor alternative explanations. We suggest that additional radial velocity and direct imaging measurements are necessary to constrain the presence of gas giants in systems with a single transiting planet.

\end{abstract}
\section{Introduction}\label{int}

\subsection{The \textit{Kepler} Dichotomy}

Since 2009, the \textit{Kepler} mission and its follow-up observations have found more than 4,000 exoplanet candidates. With an estimated fidelity of more than $90\%$ \citep[e.g.][]{morton11} the abundant \textit{Kepler} data have ushered in a new age of statistical exoplanet studies leading to a number of important discoveries about the nature of the planetary population within our galaxy \citep{lissa14}.  In general, small planets have been found to be more abundant \citep{howard12,petigura13,dres13,bata14,silb15}.  Evidence from the planet radius distribution further implies a division between rocky terrestrial planets with radii $R_{p} < 1.6 R_{\oplus}$ and somewhat larger sub-Neptunes with hydrogen/helium envelopes and $\sim 1.7 R_{\oplus} < R_{p} < 3 R_{\oplus}$ \citep{fulton17,lopez14,rogers14,wolf15,howe14}.  

One particularly impactful discovery of the Kepler mission has been the population of systems with multiple transiting planets, which typically exhibit compact short-period orbits and mutual inclination dispersions of only a few degrees \citep[e.g.][]{fabry14}.  These dynamically rich systems have allowed for masses to be determined via transit timing variations (TTVs) and for theories of planet formation to be put to the test.  Multi-transiting systems constitute $\sim 20\%$ of the \textit{Kepler} sample\footnote{Data taken from NASA's Exoplanet Archive cumulative KOI table. \url{https://exoplanetarchive.ipac.caltech.edu}.}.  While this fraction may seem large, given the low geometric probability of transit, single-component models that attempt to recover the \textit{Kepler} planetary multiplicity distribution \emph{under-predict} the number of observed single transiting systems by a factor of three \citep{lissa11,Hansen12,weissbein12}. For this reason, the general consensus is that planetary systems are divided into at least two populations: one with multiple small planets and low mutual inclinations, and the other with larger planets and either inherently lower multiplicities or high mutual inclination dispersions. \citep{moria16,lai17,johan12,ballar16} This distinction is known as the \textit{Kepler} dichotomy, and its cause is still unknown.

\subsection{Possible Causes}

Several studies that have attempted to explain the \textit{Kepler} dichotomy suggest a relationship between this phenomenon and the presence of undetected external gas giant companions in single transiting systems. Moreover, since gas giants are much more common in stars with a high metal content while smaller planets are found to exist in systems with a wider range of metallicities \citep{fischer05,johnson10,mayor11,buch12,neves13,wang15}, it is possible that a relationship between high metallicity and single transiting systems exists. 

The link between outer giant planets and single transiting systems could arise in a number of ways.  For instance, \citet{johan12} found that neither planetary collisions nor ejections can recover the observed \textit{Kepler} multiplicity in timescales commensurate with the lifetime of a typical planetary system. Instead, they concluded that the \textit{Kepler} dichotomy must be caused by gas giants starving the inner disk during planetary formation -- a scenario also proposed by \citet{moria15} and \citet{latham11}. Indeed, gas giants have also been found to act as barriers to the inward migration of the common super-Earths in planet formation simulations \citep{izid15}. 

Alternatively, the observed excess of systems with a single transiting planet could be the result of large mutual inclinations caused by the disruptive effects of gas giant companions. Analyses done by \citet{lai17} reveal that external giant planets can excite mutual inclinations in two-planet systems. Specifically, strong resonance features appear when the external planet is more massive than the innermost one -- a characteristic of many multi-planet systems \citep{weiss17}.  Building on this work, \citet{read17} found that a combination of inherently inclined systems and systems perturbed by a gas giant best predicts the abundance of single transiting planets. As a result, they propose a testable hypothesis that systems with a single transiting planet and a hidden gas giant companion, and systems with two transiting planets with high inherent mutual inclinations must exhibit different metallicity relationships.

On the other hand, a correlation with stellar metallicity might not be a necessary corollary to the physical origin of the \textit{Kepler} dichotomy. For example, \citet{lai17} mention that instead of an external gas giant, a stellar companion -- whose presence would not be associated with high metallicity -- can also excite mutual inclinations of inner planets. \citet{wang15} performed a direct imaging search for the stellar companions to \textit{Kepler} systems but did not obtain sufficient data to make claims about the occurrence of stellar companions as a function of planetary multiplicity. Furthermore, a straightforward comparison by \citet{swift} shows the metallicity distributions of cool dwarf singles and multis are indistinguishable from each other.

There are other possible indications that considerations related to the planets' host stars that might be responsible for the \textit{Kepler} dichotomy. For example, according to \citet{moria16}, the multiplicity distributions are different in GK than in M dwarf stars --- while \nicefrac{1}{3} of M dwarfs host multiple transiting planets, only $20\%$ of GK stars do. They also proposed the inferred difference in system architectures might be caused by a variety of disk surface density profiles during planetary formation. Additionally, they predict that single transiting systems should have higher stellar obliquity and larger average orbital eccentricities. In a similar manner, \citet{spalding16} found that stars with higher effective stellar temperature ($T_{eff} > 6200$ K) host considerably more single transiting planets than colder stars. Since hot stars are found to have a wide range of obliquities \citep[e.g][]{winn10, maz15}, they suggested that early on spin-orbit misalignment can produce the increased mutual inclinations responsible for the \textit{Kepler} dichotomy.  

Finally, the over-abundance of systems with only one transiting planet might not necessarily imply the existence of an architecture dichotomy. Recent simulations show that using a flat planetary disk model (instead of a flared one) removes the need for a \textit{Kepler} dichotomy \citep{bova17}. These simulations reveal that with such a disk, planets with small periods show greater mutual inclinations, ultimately making the simulated planetary systems consistent with \textit{Kepler} detections. Simulations using a single mutual inclination distribution that accounts for planetary instability are also able to match the observed multiplicity without the need for an architecture dichotomy \citep{izid17}.

\subsection{Outline of this Paper}
In this paper, we present a statistical analysis regarding the distribution of stellar metallicities in \textit{Kepler} Objects of Interest (KOIs) hosting single vs.~multiple transiting planets. Our goal, motivated by the aforementioned possible sources of the \textit{Kepler} dichotomy, is to confirm or reject the existence of a correlation between stellar metallicity ([Fe/H]) and planet multiplicity.  In the case that the presence of a gas giant companion is the main cause for the  dichotomy --- either by preventing more planets from forming or hiding them by exciting mutual inclinations --- we would expect to find higher stellar metallicities associated with single-transiting systems. While most [Fe/H] data from the Kepler Input Catalog (KIC) have high uncertainties and systematic errors \citep{dong14, brown11} we rely on the more precise stellar parameters given by the California-Kepler Survey (CKS), which accurately classify a subsample of 1305 KOIs\footnote{\url{https://github.com/California-Planet-Search/cks-website}} \citep{pet17,john17}. Additionally, we include the physical parameters of 104 M dwarf stars provided by \citet{swift}. By using high resolution and high signal-to-noise spectroscopy, the CKS data provides precise constraints on host star parameters, with a typical precision of 0.04 on [Fe/H]. In the case of the M dwarf sample, the typical precision on [Fe/H] is 0.14.

In Section~\ref{sample}, we describe the process we follow to build the sample of KOIs for our study, as well as our sub-classifications for these systems. Section~\ref{analysis} outlines our [Fe/H] distribution analysis and the statistical comparisons we carry out. In this section, we also describe a Monte Carlo probability simulation to determine an upper boundary to the fraction of single transiting systems that could additionally host a hidden gas giant. In Section~\ref{fin}, we discuss our results and move on to comment on their implications for the origin of the \textit{Kepler} dichotomy and the evolution of planetary systems in general.

\section{Sample Selection and Categories}\label{sample}

\begin{figure*}
\includegraphics[width=1\textwidth]{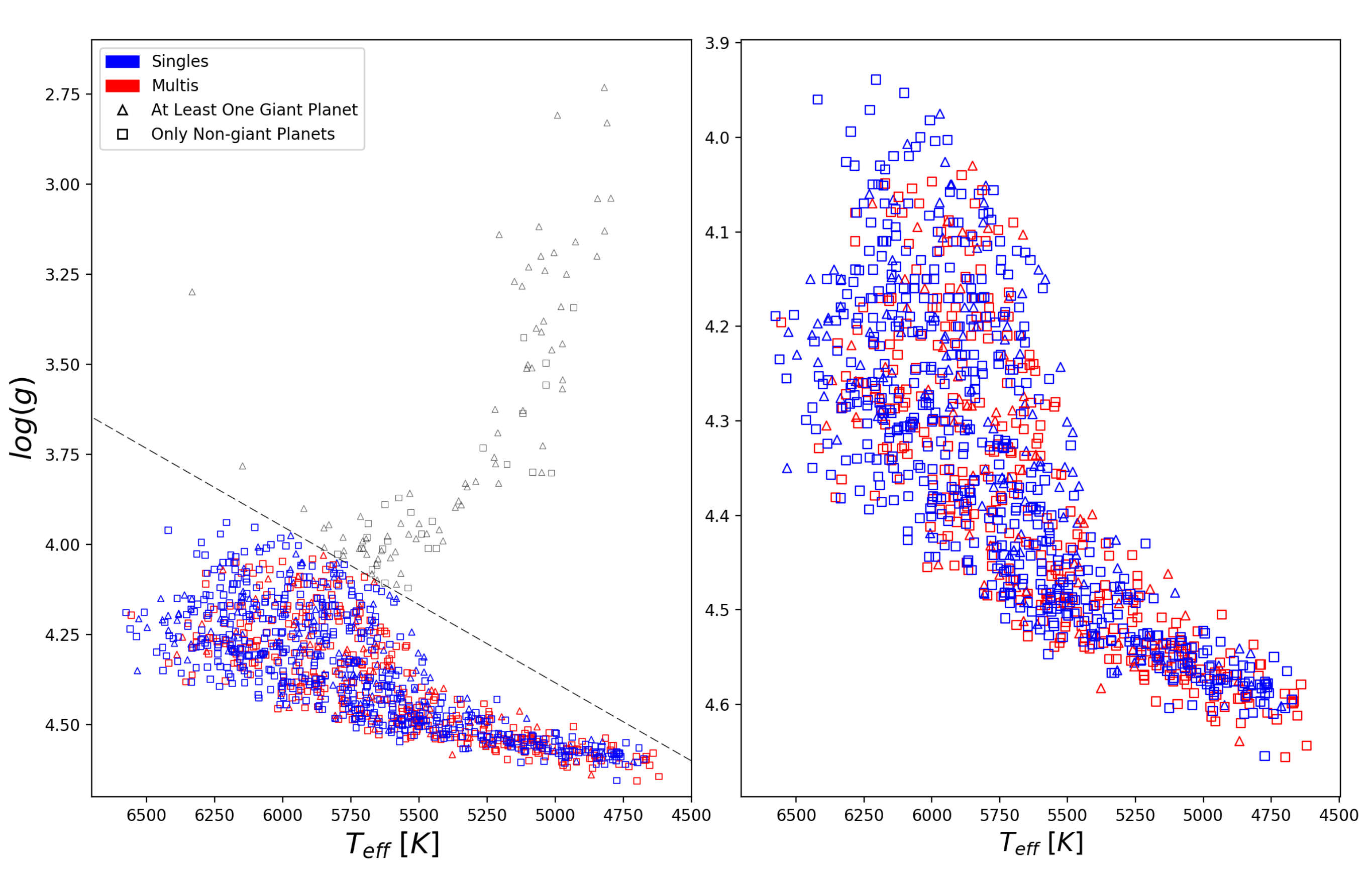}
\caption{Left panel: the full sample of CKS stars, with systems that host a single transiting planet in blue and those with multiple transiting planets shown in red. Triangles and squares denote systems that host at least one gas giant planet ($R_p > 3.9 R_{\oplus}$) and those that only host non-giant planets, respectively. We consider those stars above the dashed line with equation $log(g) = -4.3 \times 10^{-4} T_{eff}+6.55$ to be evolved giants and remove them from our sample. The dwarf star sample from \citet{swift} does not have available surface gravity values, and thus it is not shown in this figure. Right panel: the remaining main sequence stars, with same the color scheme as the left panel. Systems are well dispersed throughout the diagram in terms of both size and multiplicity with the exception that stars with $T_{eff} > 6200$ K are more likely to host \textit{singles}, as noted by \citet{spalding16}. \newline \label{fig:stars}}
\end{figure*}

To study the origin of the \textit{Kepler} dichotomy and a possible correlation with the presence of gas giants, we search for evidence of stellar metallicity trends associated with planetary multiplicity.  Our study examines the set of 1409 unique KOI host stars surveyed by the CKS and \citet{swift}. The latter sample is included to increase the number of M-dwarf hosts in our study sample so as to provide a more representative set of exoplanet hosts. The primary CKS sample ($\sim$3/4 of their total sample) is magnitude limited \citep{pet17} and should therefore not present any strong selection effects associated with metallicity or planetary multiplicity.  The final 1/4 of the CKS sample includes fainter stars that were selected based on scientific merit, including an explicit selection of multi-planet systems.  For this reason, the CKS catalog contains a higher fraction of multi-planet systems ($37\%$) compared to the full KIC ($\sim$20\%).  However, we do not expect this bias to affect our subsequent metallicity analysis --- only the relative fraction of single vs.~multi-planet systems in our study sample.   

We further cull the CKS sample in the following two ways.  (1) We narrow our selection to those stars that host at least one object not labeled as a false positive by the CKS.  (2) We limit our sample to main sequence hosts, and reject giant stars based on a cut in surface gravity ($\log g$) and effective temperature ($T_{eff}$), similar to the one used by \citet{win17} (see Figure~\ref{fig:stars}).  We perform this cut to remove evolved systems from our sample due to the unknown effects of post-main sequence evolution on planet formation and also because it is uncertain if the stellar metallicity correlation with Jovian planet occurrence applies to giant stars \citep[e.g.][]{pas07,mald13,ref15}.  In selecting for main sequence stars, and in later analyses that involve stellar metallicity and planetary radii of the CKS sample, we use system parameters derived from the CKS isochrone modeling data, which are less likely to underestimate the occurrence rate of evolved stars and the uncertainties of the stellar parameters \citep{mont15}.  In the case of the \citet{swift} sample, we do not use isochrone parameters, since these are not available. Moreover, given that the \citet{swift} sample does not include surface gravity parameters, we assume that the entire catalog consists of main sequence stars. The cuts described above result in a final sample of $1166$ stars hosting 2010 planets.

We next classify the remaining main sequence stars from our sample according to planet multiplicity and the radii of the planets they host.  We group the planets into three size classifications based on empirical evidence for divisions in planetary bulk composition.  Each planet is identified as a terrestrial, gas dwarf, or giant using the size cuts from \citet{buc14} --- 1.7 $R_{\oplus}$ and 3.9 $R_{\oplus}$, delineating the divisions between terrestrial / gas dwarf and gas dwarf / giant planets, respectively.  Similar size cuts for classifying exoplanets by their bulk composition have been proposed by a number of other authors \citep[e.g.][]{weiss14,hatzes15,chen17}.
Our detailed classification scheme is described, as follows:
\begin{itemize}

  \item \textit{Singles}:  Systems hosting exactly one transiting planet not classified as a false positive. (731 systems)
   \item \textit{Multis}: Systems hosting more than one planet not classified as a false positive. (435 systems)
   \item At least one giant planet (\textit{AOG}): Systems hosting at least one planet with $R_{p} > 3.9 R_{\oplus}$. (182 systems; 115 singles, 67 multis)
   \item Only non-giants (\textit{ONG}): Systems hosting only planets with $R_{p} \leq 3.9 R_{\oplus}$. (984 systems; 616 singles, 368 multis)
   \begin{itemize}
    \item Only terrestrial (\textit{OT}): Systems hosting only rocky planets with  $R_{p} \leq 1.7 R_{\oplus}$. (423 systems; 329 singles, 94 multis)
    \item Only gas dwarf (\textit{OGD}): Systems hosting only sub-Neptune planets with $1.7 < R_{p} \leq 3.9 R_{\oplus}$. (396 systems; 287 singles, 109 multis)
    
    \end{itemize}
\end{itemize}
Each planet size category is further subdivided into single and multiple transiting systems (\textit{singles} and \textit{multis}), as detailed above. 

\section{Metallicity Analysis} \label{analysis}

\subsection{Metallicity Distribution Statistical Tests}

\begin{figure*}[t]
\includegraphics[width=0.9\textwidth]{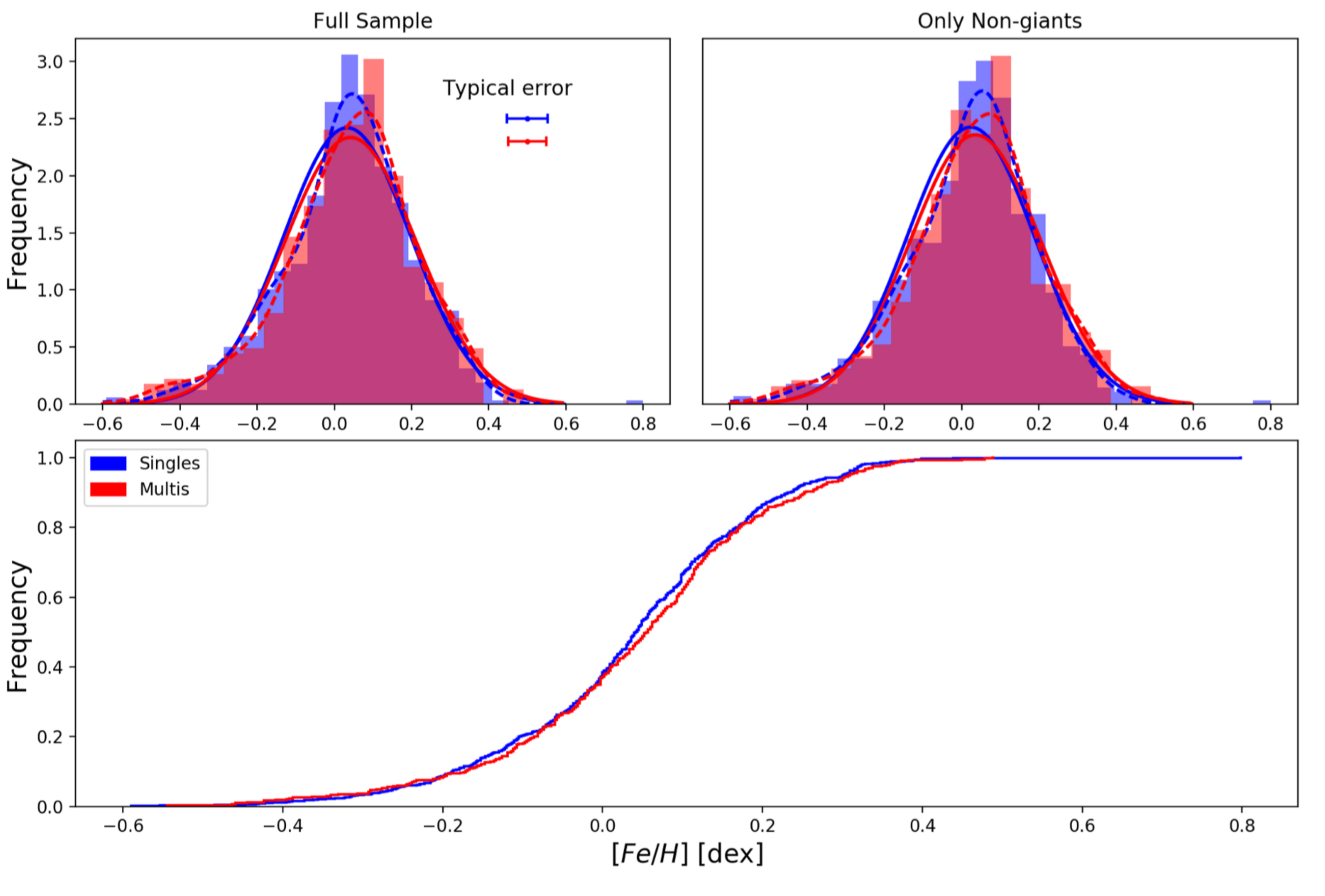}
\centering
\caption{Top left: Histograms of $[Fe/H]$ distributions for \textit{singles} (blue) and \textit{multis} (red), from the full sample of stars. Gaussian fits are shown with solid lines, and the dashed lines are kernel density estimations. Through an Anderson-Darling test for normality, we determine that the Gaussian fits are valid with $95\%$ certainty. Top right: same as top left, but only for systems with no known giant planets. Bottom: the metallicity cumulative distribution functions of singles and multis from the full sample. \label{fig:SMcomp}
\newline}
\end{figure*}

The occurrence rate of giant planets is enhanced by almost a factor of five in metal rich stars, relative to planets with $R < 3.9 R_{\oplus}$ \citep{wang15}. For this reason, if the \textit{Kepler} dichotomy is caused by gas giants (that starve the inner disk during planetary formation, excite mutual inclinations, and/or cause collisions between inner planets), we would expect singles to exhibit some correlation with high stellar metallicity. Nevertheless, a simple comparison reveals that \textit{multis} instead have a mean $[Fe/H]$ that is $0.009$ dex higher than that of singles -- a small but insignificant difference compared with the typical error on metallicity estimates from the CKS data of $\sim \pm 0.04$ (see Figure~\ref{fig:SMcomp}). 

We move on to compare the metallicity distributions in more detail. Using a two-sample Kolmogorov-Smirnov (K-S) test, we aim to determine if the underlying distribution of $[Fe/H]$ in singles differs significantly from that of multis. If the $p$ value returned by the K-S test is smaller than 0.05, or the $D$ statistic is greater than $1.36\sqrt{(N_s+N_m)/(N_sN_m)}$, where $N_s$ is the number of singles and $N_m$ the number of multis, we can conclude that the two samples are derived from different underlying distributions at greater than 95\% confidence. We perform the K-S test for singles and multis in the full sample, as well as for the various sub-samples ($AOG$, $ONG$, $OT$,  and $OGD$) to examine any differences across planet size categories. Table~\ref{table:kstab} gives our results, while the lower panel of Figure~\ref{fig:SMcomp} shows the cumulative distributions for the data from the full sample. We also perform two-sample Anderson-Darling tests on all the previous cases, which always produce values consistent with the K-S results.  In all cases, we do not find evidence to reject the null hypothesis that both populations are derived from the same metallicity distribution.  We do derive a relatively low $p$ value for the sub-Neptune sample, which we attribute to a higher average metallicity in the multi gas-dwarf systems (discussed in more detail, below).

\begin{deluxetable}{ccccc}
\tablecaption{K-S test metallicity comparison for single- and multi-transiting systems \label{table:kstab}}
\tablehead{ 
\colhead{Sample} & \colhead{$N$} & \colhead{$N_s$} & \colhead{$N_m$} & \colhead{$p$}}
\startdata 
Full Sample & $1166$ & $731$ & $435$ & $0.43$\\
Only Terrestrials & $423$ & $329$ & $94$ & $0.40$\\
Only Gas Dwarfs & $396$ & $287$ & $109$ & $0.12$\\
Only Non-giants & $984$ & $616$ & $368$ & $0.46$\\
At least one giant & $182$ & $115$ & $67$ & $0.79$\\
\enddata
\end{deluxetable}

\begin{deluxetable*}{c|c|c|c}
\tablecaption{Two-sided K-S tests between $[Fe/H]$ for the specified samples}
\tablehead{ 
\colhead{Categories} & \colhead{Full Sample} & \colhead{\textit{singles}} & \colhead{\textit{multis}}\\ & \colhead{$p$} & \colhead{$p$} & \colhead{$p$} }
\startdata
Only terrestrials - Only non-giants & $0.86$ & $0.99$ & $0.23$\\
Only terrestrials - Only gas dwarfs & $0.43$ & $0.66$ & $0.13$\\
Only terrestrials - At least one giant & $<<0.05$ & $0.01$ & $<<0.05$\\
Only gas dwarfs - Only non-giants & $0.99$ & $0.99$ & $0.87$\\
Only gas dwarfs - At least one giant & $<<0.05$ & $<<0.05$ &  $0.41$\\
Only non-giants - At least one giant & $<<0.05$ & $<<0.05$ & $0.05$\\
\enddata
\tablenotetext{}{Comparisons that yield $p<0.05$ come from different distributions at greater than 95\% confidence. \label{table:kstab2}}
\end{deluxetable*}

To further disentangle the effects of planet size and planet multiplicity, we proceed to compare the $[Fe/H]$ distribution of each of our samples with one another, first considering all systems, then looking at only single transiting systems and only multiple transiting systems separately (Table~\ref{table:kstab2}). As expected, we find the metallicity distribution for systems with only terrestrial planets is inconsistent with that of the giant planet hosts --- the latter are associated with higher metallicity.  This result holds at a comparable significance level of greater than 3-$\sigma$ for both the \textit{Kepler} singles and multis, again indicating no underlying differences in the metallicity distribution as a function of planet multiplicity in these systems.  We similarly find a high degree of inconsistency between the metallicity distribution for systems hosting giant planets and those only hosting non-giants, with one notable exception.  The $[Fe/H]$ distribution of systems that only host multiple gas dwarfs appears to be congruent with those that host at least one giant (and equivalently incongruent with those that host only terrestrial planets).  This may be a result of a greater inventory of solid material being needed to form multiple massive planetary cores for a multi gas-dwarf system.  It could also indicate that some gas dwarfs --- specifically those in multi-planet systems --- may be the eroded remains of larger giant planets.

\begin{figure}
\centering
\includegraphics[width=0.48\textwidth]{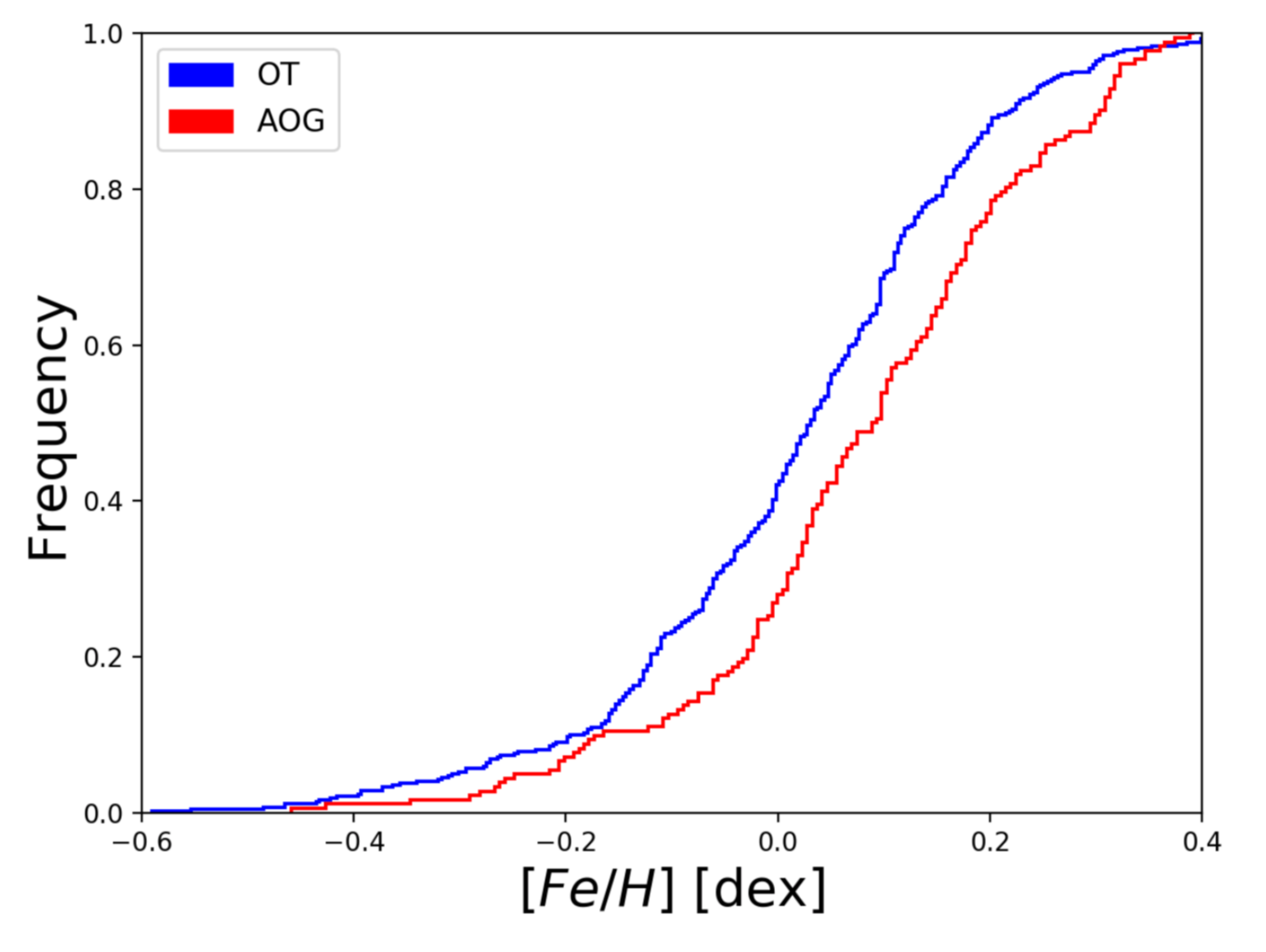}
\caption{Comparison of the cumulative distribution of $[Fe/H]$ between systems that host only terrestrial planets and systems that host at least one giant planet.  \label{fig:tvg}}
\end{figure}

Our results recover the well-known stark differences between the high metallicities associated with stars hosting giant planets and the lower metallicities of those hosting only terrestrial planets \citep[e.g.][]{buc14,wang15} (Figure~\ref{fig:tvg}).  One pertinent question that arises with respect to the giant planet metallicity correlation is whether it extends to longer orbital periods, since the suggested origin of the Kepler dichotomy relates to undetected gas giants in outer orbits.  We find that limiting our K-S test of terrestrial planet vs.~gas giant hosts to only those gas giants with longer orbital periods maintains the inconsistency in metallicity between these two populations.  However, the sample of gas giant hosts in our sample with orbital period much longer than 100 days quickly becomes vanishingly small, so, like many other studies, we cannot state conclusively that the gas giant metallicity correlation extends to planets that orbit beyond snow or ice lines.  

\begin{figure}
\centering
\includegraphics[width=0.47\textwidth]{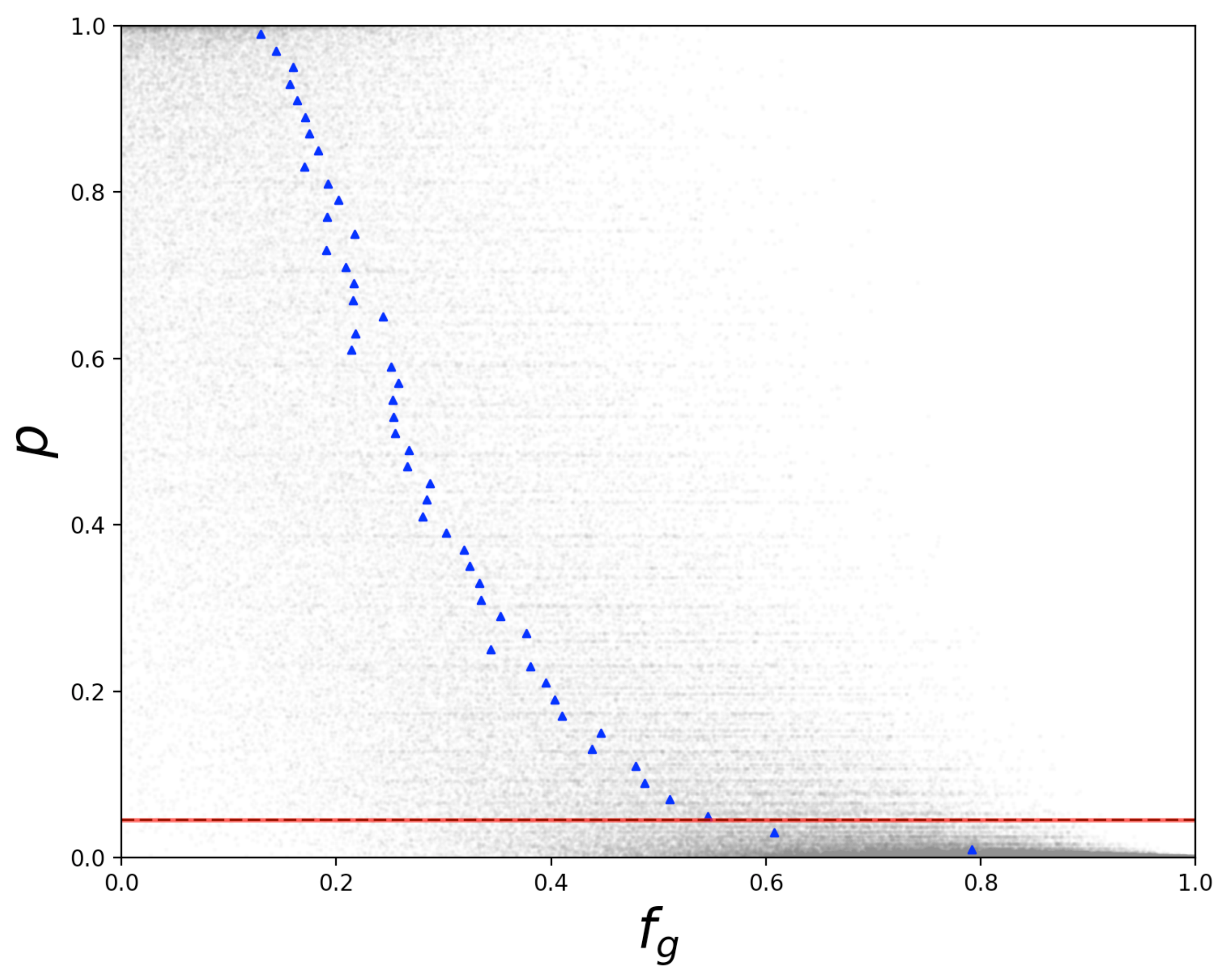}
\caption{Monte Carlo simulation results. Each gray dot represents one of $10^{5}$ realizations of a K-S test between a randomly generated sample of $ONG$ \textit{singles} and $AOG$ systems compared to the full $ONG$ \textit{singles} sample.  The blue triangles show the mean $f_g$ for $p$ values increasing in intervals of 0.02. The red line at $p = 0.0455$ is the $2\sigma$ upper boundary, which corresponds to a $\sim 55\%$ fraction of giant planet hosts that could be embedded within the $ONG$ \textit{singles} sample. \label{fig:montecarlo}}
\end{figure}

\subsection{Monte Carlo Test for Overlap Between Single Planet and Gas Giant Populations}

The lack of a metallicity increase associated with the \textit{Kepler} singles indicates that gas giants are unlikely to be the main factor behind the \textit{Kepler} dichotomy. From the 731 systems that constitute the main sequence singles population, 616 are found to not host any giant planets. Nonetheless, it is possible that, in accordance with some predictions \citep[e.g.][]{moria15,lai17,read17}, a fraction of these systems host long-period giant planets that are yet to be discovered or are hidden due to high mutual orbital inclinations.  Therefore, while there is not a substantial overlap between the giant planet host population and \textit{Kepler} singles, based on our metallicity analysis, a \emph{partial} overlap between the single transiting systems and stars hosting unseen giant planets is still possible.  For this reason, we aim to determine the maximum fraction of systems with a single non-giant transiting planet that could also host a hidden gas or ice giant. 

We run a Monte Carlo simulation of two-sided K-S tests that compare the $[Fe/H]$ distribution of a random sample with that of the giant planet hosts. First, our simulation generates a random number $0\leq f_g \leq1$, which we use to build a 182-element (the number of systems in the $AOG$ sample) array that randomly takes $[Fe/H]$ values from the $AOG$ and $ONG$ \textit{singles} populations, with replacement. For instance, if $f_g = 0$, our random sample consists of 182 random values taken only from the $ONG$ \textit{singles} sample; on the other hand, if $f_g = 1$, the random sample contains metallicities from the $AOG$ population only. Then we perform a K-S test between the randomly selected sample and the $ONG$ \textit{singles} population, and find the resulting $p$ value.  We repeat the Monte Carlo simulation $10^5$ times and show the result of each iteration (gray dots) in Figure \ref{fig:montecarlo}. The $2\sigma$ upper bound (corresponding to $p=0.0455$) is found to correspond to a value of $f_g \approx 0.55$, such that up to $\sim 55\%$ of $ONG$ \textit{singles} could host a hidden gas giant planet without altering our results from above.

\section{Discussion and Conclusions} \label{fin}

This paper has presented a statistical comparison of \textit{Kepler} planetary hosts of single and multiple transiting systems as a means of uncovering the source of the \textit{Kepler} dichotomy.  We have specifically searched for indications that \textit{Kepler} singles may be preferentially accompanied by gas giant planets, which would be marked by a signpost of higher host star metallicity in these systems.  We find no such metallicity trends and no significant differences between the metallicity distribution of \textit{Kepler} singles and multis.  This result holds up across the full main-sequence samples, as well as with the other planet-size based sub-samples that we examined.  

We recover strong differences in the metallicities of giant planet hosts vs.~non-giant hosts. Giant planets in our sample preferentially orbit metal-rich stars, whereas terrestrial planets and single gas-dwarfs do not show as strong a bias toward high metallicity host stars.  However, we also report a higher average metallicity in systems with multiple gas-dwarf planets.  These results are in line with previous findings from the literature. \citep{wang15}

Our Monte Carlo simulation shows that up to $55\%$ of systems with a single non-giant transiting planet could host an undiscovered giant, based on metallicity considerations. This finding supports a conclusion that giant planets are not the main source of the over-abundance of \textit{Kepler} singles, although they could be one contributing factor. However, since gas giants with long periods could exhibit a different metallicity correlation, a definitive answer to the cause of the \textit{Kepler} dichotomy requires supplementary analyses. Radial velocity and imaging surveys designed specifically to search for long-period giant planets in single vs.~multiple transiting systems will ultimately reveal the role of gas giants in sculpting the multiplicity distribution of close-in transiting exoplanets.

\acknowledgements
We acknowledge funding from the Research Corporation through their Cottrell College Science Awards grant program. We express our gratitude to the anonymous referee and the AAS statistical editor for their useful comments and the attention they gave to this paper. We thank Grinnell College for providing the Directed Research fellowship and resources that supported the work of CEMR. 

\bibliography{Kepler}

\end{document}